\documentclass[english,aps,pre,twocolumn]{revtex4-1}
\usepackage[T1]{fontenc}
\usepackage{amsmath}
\usepackage{amssymb}
\usepackage{graphicx}
\usepackage[most]{tcolorbox}
\usepackage{subfigure}

\begin{document}
	\title{Josephson effect in superconductor-normal dot-superconductor junctions driven out of equilibrium by quasiparticle injection}
	\author{Sarath Sankar, Julia S. Meyer, and Manuel Houzet}
	\affiliation{Univ.~Grenoble Alpes, CEA, Grenoble INP, IRIG, Pheliqs, F-38000 Grenoble, France}
	
	\begin{abstract}
		We study theoretically the {\em large} variations of the supercurrent through a normal dot that are induced by a {\em small} quasiparticle injection current from normal leads connected to the dot. We find that the supercurrent decomposes into a subgap contribution, which depends on the voltages applied to the normal leads, as well as a contribution with opposite sign from energies outside the gap, which is insensitive to the voltages. As the voltages gradually suppress the subgap contribution, a critical voltage exists above which the contribution from energies outside the gap dominates, leading to a sign reversal of the current-phase relation, namely a transition to a so-called $\pi$-junction behavior. We determine the critical voltage and analyze the robustness of the effect with respect to temperature and inelastic relaxation in the dot. \end{abstract}
	
	\maketitle
	\date{\today}
	
	\section{Introduction}

The fascinating properties of superconducting devices are in large part associated with the presence of an excitation gap for quasiparticles such that, at low temperatures, the system can be described in terms of a macroscopic variable, the superconducting phase, while being protected from dissipation. It turns out, however, that quasiparticles still play an important role in many situations. Nonequilibrium quasiparticles may be present due to the breaking of Cooper pairs, e.g., by the absorption of stray photons or cosmic rays.  They have been shown to be very difficult to get rid off and to be detrimental for  the coherence of superconducting qubits (see, e.g., Ref.~\cite{glazman2021bogoliubov} for a recent review).
On the other hand, trapped quasiparticles in the Andreev bound states that form in Josephson junctions can lead to interesting novel phenomena, such as the realization of an Andreev spin qubit~\cite{Padurariu2010,Hays2021}.

Recent experiments on gate control of the supercurrent in metallic Josephson junctions~\cite{Simoni2018,Paolucci2018,Simoni2019,Paolucci2019,Rocci2020}   have revived the interest in a better understanding of the role of quasiparticle injection from normal parts of the circuit~\cite{Golokolenov2021,Ritter2021a,Ritter2021b,Basset2021}. 
Deliberate quasiparticle injection via voltage-biased normal leads has indeed been studied earlier and shown to have important effects on the supercurrent~\cite{chang1997control,morpurgo1998hot,savin2004cold}. The possible reversal of the sign of the supercurrent upon increasing the voltage of a normal lead directly coupled to the junction was first shown, though not emphasized, theoretically in a long ballistic junction~\cite{van1991transmission}.  The resulting realization of a so-called nonequilibrium $\pi$-junction was pointed out in Ref.~\cite{volkov1995new}, where  a simpler setup consisting of a normal dot connected to two superconductors was investigated. Experimentally, a nonequilibrium $\pi$-junction was first realized in a long diffusive junction~\cite{baselmans1999reversing}.
Also subsequent work, both theoretical~\cite{wilhelm1998mesoscopic,yip1998,heikkila2002supercurrent} and experimental~\cite{huang2002observation,baselmans2002direct}, concentrated on extended junctions. (A short ballistic junction was addressed in Ref.~\cite{heikkila2002supercurrent}. However, in that case, the effect is absent.) More complicated geometries, sometimes called Andreev interferometers, have been studied as well~\cite{samuelsson2000nonequilibrium,zaikin2019sr}, but all in the long-junction limit. We note in passing that {\em equilibrium} $\pi$-junctions may be realized in superconductor-ferromagnet-superconductor junctions~\cite{Bulaevskii1977,Buzdin1982,Ryazanov2001,Kontos2002}. 

Here we revisit the simple superconductor-normal dot-superconductor setup and analyze the supercurrent in detail. Our main findings are that the nonequilibrium $\pi$-transition exists irrespective of the coupling strengths between the dot and the superconductors, and that it is robust with respect to temperature and inelastic relaxation due to electron-electron interactions. Furthermore, we show that the same phenomenon also occurs in multiterminal Josephson junctions~\cite{Cohen2018,Draelos2019,pankratova2020multiterminal,Graziano2020} that are currently under intensive investigation, as they may be used for Majorana braiding~\cite{Alicea2011,Aasen2016} and have interesting topological properties~\cite{VanHeck2014,Yokoyama2015,Riwar2016}

	Using the quasiclassical Usadel equations in the Keldysh formulation~\cite{belzig1999quasiclassical}, we study a chaotic normal dot coupled to various superconducting and normal leads. We obtain explicit analytical expressions for the currents flowing from the dot to the different leads. Whereas the links to the normal leads only carry dissipative currents, the links to the superconducting leads may carry both dissipative currents and dissipationless supercurrents. We find that the dissipative and supercurrents can be distinguished by their behavior under a gobal change of sign of all the voltages: while the dissipative currents are odd, the supercurrents are even under such a sign reversal. 
	
The distinct behavior under a global change of sign of all voltages allows us to separately study the supercurrents, which can be expressed as an energy-integral over the product of a spectral function and a distribution function. While the spectral function only depends on the spectral properties of the leads and the couplings between the dot and the leads, the distribution function also depends on the distribution functions of the leads. Interestingly the contributions to the supercurrent from subgap energies and from energies outside the gap show a quite different behavior.  The subgap contributions depend only on the voltage-dependent distribution functions of the normal leads, which is easy to understand as the gap in the superconducting leads prevents thermalization at these energies. By contrast, the contributions from energies outside the gap depend only on the equilibrium distribution functions of the superconducting leads. Thus the quasiparticle injection from the normal leads only affects the subgap contributions and, in particular, suppresses them upon increasing the bias voltages. 

Furthermore, we show that, at fixed superconducting phases, the signs of the subgap contributions and the contributions from energies outside the gap are opposite, such that the two contributions are in competition. This effect can be traced back to the negative sign of the spectral current function at energies above the gap, as emphasized in the two-terminal case in Ref.~\cite{Levchenko2006}. In equilibrium, we find that the subgap contributions always dominate and, therefore, fix the signs of the supercurrents. However, as the applied voltages to the normal leads are increased, the gradual suppression of the subgap contributions leads to a supercurrent sign reversal at a critical value of the voltages, where the contributions from energies outside the gap become dominant. 

In a two-terminal junction, a sign reversal of the supercurrent corresponds to a transition from a conventional 
junction to a so-called nonequilibrium $\pi$-junction. We study this effect in detail in the case of weak and strong coupling of the dot to the superconducting leads, both at zero temperature and close to the superconducting critical temperature, $T_c$. 
At zero temperature, we find that the transition always occurs at voltages $V^*<\Delta_0/e$, where $\Delta_0$ is the zero-temperature superconducting gap. Upon further increasing the voltage  above $\Delta_0/e$, the critical current saturates 
to a value that is smaller than or comparable to the equilibrium critical current. Close to $T_c$, 
the behavior is qualitatively different. At strong coupling, the transition is pushed up to voltages $V^*>\Delta_0/e$. By contrast, at weak coupling, the finite temperature reduces the value of the critical voltage $V^*$. Interestingly, in that case, the critical current at large voltages in the $\pi$-junction regime is parametrically larger than the equilibrium supercurrent.
	
As in any nonequilibrium phenomenon, relaxation plays an important role. We study the robustness of the effects discussed above with respect to internal inelastic relaxation inside the dot due to electron-electron interactions. Modeling this inelastic relaxation by a fictitious fermionic bath~\cite{Buttiker1986}, we find that the transition survives even when the inelastic relaxation rate $\Gamma_b$ is fairly large, though it is pushed to large voltages. Thus the effect is robust.

		The  paper is organized as follows: In Sec.~\ref{sec:model}, we define the model and provide the general expressions for the currents in the quasiclassical-Keldysh formalism. In Sec.~\ref{sec:results}, we analyze the supercurrents in the absence of inelastic relaxation and obtain the critical voltage at which the $\pi$-transition happens.    
In Sec.~\ref{sec:inelrel}, the effects of inelastic relaxation within the dot due to electron-electron interactions are considered. Finally we conclude in Sec.~\ref{sec:concl}. Some details and generalizations of the setup can be found in the Appendices. 
	
	\section{The model \label{sec:model}}

	We consider a normal dot that is coupled to  normal (N) and superconducting (S) leads through tunnel barriers.  In the quasiclassical-Keldysh formalism,  the electric current through the junction connecting the dot to lead $p$ can be expressed in terms of the quasiclassical Green functions of the dot ($\check{g}$) and the lead ($\check{g}_p$), subject to the normalization condition $\check{g}_{(p)}^2=1$, as follows:
	\begin{equation}
		\label{lcurform}
		I_{p}=\frac{ G_{p}}{16e}\int_{-\infty}^\infty dE\,\tilde{I}_{p}(E)
		\end{equation}
with 
\begin{equation}
\tilde{I}_{p}=\text{Tr} \left[\hat{\tau}_{3}\left(\hat{g}_{p}^{K}\hat{g}^{A}-\hat{g}^{R}\hat{g}_{p}^{K}-\hat{g}^{K}\hat{g}_{p}^{A}+\hat{g}_{p}^{R}\hat{g}^{K}\right)\right],
	\end{equation}
	where $G_p$ is the tunnel conductance in the normal state, and $\hat{g}_{(p)}^{R/A/K}$ denotes the retarded, advanced, and Keldysh components of the quasiclassical Green function, respectively:
			\begin{equation}
		\check{g}_{(p)}=\left(\begin{array}{cc}
			\hat{g}_{(p)}^R&\hat{g}_{(p)}^K\\
			0&\hat{g}_{(p)}^A
		\end{array}\right).
	\end{equation}
The components $\hat{g}_{(p)}^{R/A/K}$ are $2\times 2$ matrices in Nambu space, and $\hat{\tau}_{1,2,3}$ are Pauli matrices acting in that space. 

If we neglect inelastic relaxation in the dot, the dot Green function is determined by the equation
	\begin{equation}
		\label{eq:green}
		[\check{h},\check{g}]=0 \quad{\rm with}\quad \check{h}=E\check{\tau}_3+i\sum_p\Gamma_p\check{g}_p.
	\end{equation}
Here $\check{\tau}_3=\hat{\sigma}_{0}\otimes\hat{\tau}_{3}$ with $\hat{\sigma}_{0}$ being the identity matrix in Keldysh space, $E$ is the energy measured from the Fermi level in the S leads, $\Gamma_p=G_p\delta/(2\pi G_Q)$, where $\delta$ is the mean level spacing in the dot and $G_Q=e^{2}/\pi$ the conductance quantum  (in units where $\hbar=1$), are the partial level widths due to the tunnel coupling of the dot to the leads.
	
The spectral Green functions $\hat{g}_{(p)}^{R/A}$ can be generally expressed as
	\begin{equation}
		\label{eq:gr_general}
		\hat{g}^R_{(p)}=\sin\theta_{(p)} \left[\sin\chi_{(p)}\hat{\tau}_1+\cos\chi_{(p)}\hat{\tau}_2\right]+\cos\theta_{(p)}\hat{\tau}_3
	\end{equation}
and $\hat{g}^{A}_{(p)}=-\hat{\tau}_3\hat{g}^{R\,\dagger}_{(p)}\hat{\tau}_3$	with complex angles $\theta_{(p)}$ and $\chi_{(p)}$.
	
For a normal lead, $\theta_p=0$ such that $\hat{g}_{p}^{R/A}=\pm\hat\tau_3$. Assuming that all the superconducting leads are grounded and have the same gap amplitude and different superconducting phases $\phi_p$, $\Delta_p\equiv\Delta e^{i\phi_p}$, their Green functions are obtained with 
	\begin{equation}
		\label{eq:sclead_gr}
		\chi_p = \phi_p	 \quad \text{and}\quad \theta_p = \theta_S, \quad\text{where} \quad \tan \theta_S=\frac{i\Delta}{E+i0^+}.
	\end{equation}
The Green functions  $\hat{g}^{R/A}$ of the dot are determined using the respective blocks of Eq.~\eqref{eq:green}, $[\hat{h}^{R/A},\hat{g}^{R/A}]=0$ with $\hat{h}^{R/A}=E\hat{\tau}_3+i\sum_p\Gamma_p\hat{g}^{R/A}_p$. As $(\hat{h}^{R/A})^2$ is proportional to the identity matrix, it is easy to see that $\hat{g}^{R(A)}=\hat{h}^{R(A)}/\xi^{R(A)}$  with $\xi^{R(A)}=\pm\text{sign}(E)\sqrt{(\hat{h}^{R(A)})^{2}}$,  where the sign convention is chosen to match the normal state result $\hat{g}^{R(A)}=\pm\hat{\tau}_{3}$ in the limit of vanishing coupling to the leads. Note that $\xi^{A}=-(\xi^{R})^*$. Thus the dot parameters $\theta$ and $\chi$ are given as
	\begin{equation}
		\label{eq:theta_chi}
		\sin\theta=\frac{i\Gamma_\phi\sin\theta_S}{\xi^R},\quad \sin\chi=\frac{\sum_{s} \Gamma_{s}\sin \phi_{s}}{\Gamma_\phi} 	\end{equation}
		with $\Gamma_{\phi}=\sqrt{\sum_{s,s'}\Gamma_{s}\Gamma_{s'}\cos\phi_{ss'}}$ and
	\begin{equation}
		\label{eq:gmphi_xir}
		\xi^{R}=\text{sign}(E)\sqrt{(E+i\Gamma_N+i\Gamma_S\cos\theta_S)^{2}-\Gamma_{\phi}^2\sin^{2}\theta_S}\, ,
	\end{equation}
	where the sum over leads $s$ is restricted to the superconducting leads,  $\phi_{ss'}=\phi_{s}-\phi_{s'}$, and $\Gamma_N$ and $\Gamma_S$ are the sums of the partial level widths associated with the N and S leads, respectively. Note that $\chi$ is real.

	The Keldysh part of the equilibrium Green function  reads $\hat{g}^{K}_{p}=(\hat{g}^{R}_{p}-\hat{g}^{A}_{p})f_{L0}(E)$, where $f_{L0}(E)=\tanh (E/{2T})$ (in units where $k_B=1$). For a normal lead $n$ biased at voltage $V_{n}$, one obtains $\hat{g}^{K}_{n}(E)=2 \hat{\tau}_3[1-2f(E-eV_{n}\hat{\tau}_3)]$, where $f(E)$ is the Fermi-Dirac distribution function. It can also be written in the form $\hat{g}^{K}_{n}(E)=2\hat{\tau}_3 f_{Ln}(E)+2\hat{\tau}_0 f_{Tn}(E)$, where the longitudinal component of the distribution function,  $f_{Ln}(E)=[f(-E-eV_{n})-f(E-eV_{n})]$, is odd in energy and even in voltage, whereas the transversal component of the distribution function, $f_{Tn}(E)=1-f(E-eV_{n})-f(-E-eV_{n})$, is even in energy and odd in voltage. 
	
To obtain the Keldysh part of the dot Green function, following Refs.~\cite{snyman2009bistability,catelani2010effect}, we combine the Keldysh component of the normalization condition, $\hat{g}^{R}\hat{g}^{K}+\hat{g}^{K}\hat{g}^{A}=0$, and the Keldysh component of Eq.~\eqref{eq:green} to find
	\begin{equation}
		\label{eq:Keldysh}
		\hat{g}^{K}=\frac1{\xi^R+\xi^A}(\hat{h}^{K}-\hat{g}^{R}\hat{h}^{K}\hat{g}^{A}).
	\end{equation} 
Using
	\begin{eqnarray}
	\hat h^K&=&i\sum_p\Gamma_p\hat g_p^K\\\
	&=&(\hat h^R-\hat h^A)f_{L0}+2i\sum_{n}\Gamma_{n}\left[\hat\tau_3(f_{Ln}-f_{L0})+\hat\tau_0f_{Tn}\right],
	\nonumber \end{eqnarray} 
where the sum over leads $n$ is restricted to the normal leads, and the identities derived in Appendix~\ref{ap_relation}, it can be written in the form
	\begin{eqnarray}
		\label{eq:Keldysh3}
		\hat{g}^{K}=(\hat{g}^{R}-\hat{g}^{A})f_L+(\hat{g}^{R}\hat{\tau}_3-\hat{\tau}_3\hat{g}^{A})f_T
	\end{eqnarray} 
	with 
	\begin{eqnarray}
		f_L&=&f_{L0}+\frac {C}{{\rm Im}\,\xi^R}\sum_{n}\Gamma_{n}(f_{Ln}-f_{L0}),\label{eq:fl}\\
		f_T&=&\frac {1}{C\,{\rm Im}\,\xi^R}\sum_{n}\Gamma_{n}f_{Tn},\label{eq:ft}
	\end{eqnarray}
where
	\begin{equation}
		\label{eq:id2}
		C=\frac{1+\cos\theta\cos\theta^*+\sin\theta\sin \theta^*}{\cos\theta+\cos\theta^*}.
	\end{equation}
Note that for $|E|<\Delta$, $C$ takes a particularly simple form (see Appendix \ref{ap_relation}), $C={\rm Im}\,\xi^R/\Gamma_N$, such that the subgap distribution function only depends on the distribution functions of the normal leads,
	\begin{equation}
		\label{eq:fL_simple}
		f_L(E)=\frac{1}{\Gamma_N}\sum_{n}\Gamma_{n}f_{Ln}(E)\quad{\rm if}\quad |E|<\Delta.
	\end{equation}
This reflects the fact that no thermalization with the superconducting leads is possible at these energies. At energies $|E|\gg\Delta$, one recovers the normal state result, where $f_L=\sum_p\Gamma_pf_{Lp}$ is a weighted sum of the distribution functions of all the leads.

We now have all the elements necessary to evaluate the currents as given by Eq.~\eqref{lcurform}.
The current to a normal lead takes the form
	\begin{equation}
		I_{n}=\frac{G_{n}}{4e}\int dE\,\tilde{I}_{n}(E) \quad {\rm with} \quad \tilde{I}_{n}=2(f_{T}-f_{Tn}){\rm Re}(\cos\theta). \label{eq:cur_N}
	\end{equation}
It is a dissipative current that is odd under flipping the signs of all the voltages (as can be readily deduced from the expression for $f_{Tn})$.

The current to a superconducting lead can be decomposed  into a  dissipative current $I_{s}^{dis}$ and a dissipationless supercurrent $I_{s}^{S}$ \cite{likharev1979,barone1982,volkov1995new}, namely,

	\begin{eqnarray}
		I_{s}&=&I_{s}^{S}+I_{s}^{dis}=\frac{G_{s}}{2e}\int dE\,\left(\tilde{I}_{s}^{S}(E)+\tilde{I}_{s}^{dis}(E)\right)\label{eq:cur_S_tot}
		\end{eqnarray}
		with
		\begin{widetext}
		\begin{eqnarray}
		\tilde{I}_{s}^{S}(E)&= &\frac{\sum_{s'}\Gamma_{s'}\sin\phi_{s's}}{\Gamma_{\phi}} \left[f_L{\rm Im}(\sin\theta){\rm Re}(\sin\theta_S)+f_{L0}{\rm Re}(\sin\theta){\rm Im}(\sin\theta_S)\right],\label{eq:cur_S_sup}\\
		\tilde{I}_{s}^{dis}(E)&=&f_T\left[{\rm Re}(\cos\theta){\rm Re}(\cos\theta_{S})-\frac{\sum_{s'}\Gamma_{s'}\cos\phi_{s's}}{\Gamma_{\phi}}{\rm Re}(\sin\theta){\rm Re}(\sin\theta_{S})\right].\label{eq:cur_S_qp}
	\end{eqnarray}
			\end{widetext}
Under a global flip in the sign of the voltages in the normal leads we see that $f_L$ is even where as $f_T$ is odd which results in the supercurrent being even and the dissipative current being odd. Thus the dissipative currents and supercurrents are conserved separately:
	\begin{equation}
		 \sum_{n}I_{n}+\sum_{s}I_{s}^{dis}=0\quad {\rm and} \quad\sum_{s}I_{s}^{S}=0.
	\end{equation}
We further note that the two components in our setup satisfy an additional symmetry property: the supercurrents and disippative currents are odd and even, respectively, under a global sign flip of the superconducting phases. (This result may not survive in the case of a finite-length normal region~\cite{dolgirev2019phase}.)
	
Here we are interested in the supercurrents $I_{s}^S$. 
As $\sin\theta_{S}$ is purely real for $|E|<\Delta$ and purely imaginary for $|E|>\Delta$ (see Eq.~\eqref{eq:sclead_gr}), we can distinguish two contributions in Eq.~\eqref{eq:cur_S_sup}: a subgap contribution that depends on the voltage-dependent dot distribution function $f_L$, and a contribution from energies outside the gap that depends on the equilibrium distribution function $f_{L0}$. 
Thus, using Eq.~\eqref{eq:fL_simple} and the observation that the integrand is even in energy,
\begin{widetext}
	\begin{equation}
			\label{supcur_gen_exp-v0}
				I_{s}^{S}=\frac{\sum_{s'}G_{s}\Gamma_{s'}\sin\phi_{s's}}{e\Gamma_{\phi}}\left\{\sum_{n}\frac{\Gamma_{n}}{\Gamma_N}\int_0^\Delta dE\,f_{Ln}{\rm Im}(\sin\theta)\sin\theta_S -i\int_\Delta^\infty dE\, f_{L0}{\rm Re}(\sin\theta)\sin\theta_S\right\}.
	\end{equation}
	\end{widetext}
The contribution from energies outside the gap vanishes when the dot is perfectly coupled to the superconductors ($\Gamma_S\to\infty$) such that a short ballistic junction is formed. In that case, the supercurrent is carried by Andreev states only. A finite value of $\Gamma_S$ has a similar effect as a non-zero length or a finite Thouless energy, yielding a continuum contribution to the supercurrent~\cite{Kulik1970,Levchenko2006}. The presence of this continuum contribution is crucial for the phenomena described here.

Equation~\eqref{supcur_gen_exp-v0} can be further simplified using Eqs.~\eqref{eq:theta_chi} and \eqref{eq:gmphi_xir}, namely
\begin{widetext}
	\begin{equation}
		\label{supcur_gen_exp}
		I_{s}^{S}=\frac1e\sum_{s'}G_{s}\Gamma_{s'}\sin\phi_{s's} \left\{\sum_{n}\frac{\Gamma_{n}}{\Gamma_N}\int_0^\Delta dE\,f_{Ln}(E)j
	(E)+\int_\Delta^\infty dE\, f_{L0}(E)j(E)\right\}\nonumber
	\end{equation}
\end{widetext}
with 
\begin{equation}
j(E)=\frac{{\rm Re}\,\xi^R}{|\xi^R|^2}\sin^2\theta_S=\frac{{\rm Re}\,\xi^R}{|\xi^R|^2}\frac{\Delta^2}{\Delta^2-E^2}.
\end{equation}
(We will see in the next section that the singularity at $|E|=\Delta$ is integrable.)

As ${\rm Re}\,\xi^R(E>0)$ is positive, we make the important observation that the supercurrent,
	\begin{equation}
		\label{supcur_gen_exp}
		I_{s}^{S}=\frac1e\sum_{s'}G_{s}\Gamma_{s'}\sin\phi_{s's}\left\{\sum_{n}\frac{\Gamma_{n}}{\Gamma_N}K_{n}^<+K^>\right\},
	\end{equation}
results from a competition between positive subgap contributions determined by 
\begin{equation}
		\label{klg1}
		K_{n}^{<}=\int_{0}^{\Delta}dE\;f_{Ln}(E)j(E) >0
\end{equation}
and negative contributions from energies outside the gap determined by
\begin{equation}
		\label{klg2}
		 K^{>}=\int_{\Delta}^{\infty}dE\;f_{L0}(E)j(E)<0.
		\end{equation}
As we will see in the following, it is this competition that leads to a nonequilibrium $\pi$-junction. In particular, we find that, in equilibrium, when $f_{Ln}=f_{L0}$, the supercurrent is dominated by the positive subgap contributions $K_{n}^<$. Out of equilibrium, the modified distribution functions $f_{Ln}$ suppress the subgap contributions such that, eventually, the supercurrent will be dominated by the negative contribution $K^>$ from energies outside of the gap. The resulting sign change signals the transition to a $\pi$-junction behavior. In the following section, we determine the supercurrent as a function of the applied voltages in various regimes.

	\section{Voltage-dependent supercurrent in the absence of inelastic relaxation \label{sec:results}}

For the main part of this paper, we will consider a specific setup with two superconducting leads, phase-biased at a phase difference $\phi_2-\phi_1=\phi$, and normal leads, voltage-biased at voltages  with the same absolute value, $|V_n|=V$. As $f_{Ln}$ is an even function of the voltage, the distribution functions of all the normal leads are the same, such that all the $K_n$ in Eq.~\eqref{supcur_gen_exp} are the same, and the expression for the supercurrent simplifies to
	\begin{equation}
		\label{supcur_gen_exp}
		I_{1}^{S}=-I_2^S=\frac{\Gamma_S}{eR}\sin\phi(K^>+K^<),
	\end{equation}
where $K^<$ is given by Eq.~\eqref{klg1} and $R=(G_1+G_2)/G_1G_2$. This formula allows us to numerically evaluate the supercurrent in all parameter regimes. In the following, to get a better understanding of the results, we discuss limiting cases where an analytical expression for the supercurrent can be obtained. Note that the dissipative currents depend on the signs of the voltages in the normal leads.  The dissipative currents in the superconducting leads vanish in a symmetric setup in which $\Gamma_{N>}=\Gamma_{N<}$, where $\Gamma_{N>} (\Gamma_{N<})$ is the sum over the $\Gamma_n$ of the leads biased at $+V (-V)$. This follows from the separate conservation of supercurrents and dissipative currents. In an asymmetric setup, the dissipative currents would contribute to the measured critical current \cite{kutchinsky1999,seviour2000,bezuglyi2003}.  However,  in the regime, where the normal leads that drive the system out of equilibrium are weakly coupled to the dot, $\Gamma_N\ll\Gamma_S$, this effect is negligible. In the following, we will concentrate on that regime.

While a non-vanishing coupling is necessary to establish the  out-of-equilibrium distribution function $f_L$, we can see from Eqs.~\eqref{eq:fL_simple} that $f_L(E<\Delta)$, which enters the expression for the supercurrent, does not depend on the absolute magnitude of the couplings. Namely, the value of $\Gamma_N$ affects the supercurrent, Eq.~\eqref{supcur_gen_exp}, only via the spectral current $j(E)$. As $j(E)$ is non-vanishing in the absence of a coupling to the normal leads, we may evaluate it at $\Gamma_N=0$ to obtain the result in leading order.
Then the expression for $\xi^R$  determining $j(E)$ takes the form
	\begin{equation}	
		\xi^{R}=
		\begin{cases}
			\sqrt{\left(E+\frac{\Gamma_SE}{\sqrt{\Delta^2-E^2}}\right)^2-\frac{\Gamma_{\phi}^2\Delta^2}{\Delta^2-E^2}},\quad E<\Delta,\\
			\, \\
			\sqrt{\left(E+i\frac{\Gamma_SE}{\sqrt{E^2-\Delta^2}}\right)^2+\frac{\Gamma_{\phi}^2\Delta^2}{E^2-\Delta^2}},\quad E>\Delta.
		\end{cases}
	\end{equation}
	Note that  $\xi^R$ vanishes for some $E=E_g$, where $E_g$ satisfies the equation,
	\begin{equation}
		\label{eq-Eg}
		E_g=\frac{\Gamma_\phi \Delta}{\Gamma_S+\sqrt{\Delta^2-E_g^2}}.
	\end{equation}
	$\xi^R$ is purely imaginary for $E<E_g$, and consequently $j(E<E_g)$ vanishes. Furthermore, in the interval, $E_g<E<\Delta$, $\xi^R$ is real such that the spectral current simplifies to $j(E)=\{\xi_R[1-(E/\Delta)^2]\}^{-1}$.

To proceed further we will	study two limiting cases: weak coupling $\Gamma_S \ll \Delta_0$ in Sec.~\ref{sec-weak} and strong coupling $\Gamma_S \gg \Delta_0$ in Sec.~\ref{sec-strong}. We start by considering the zero-temperature case, where $f_{L0}(E>0)=1$ and $f_{L}(E>0)=\Theta(E-eV)$. We determine the  critical current as a function of voltage and, in particular, determine the voltage $V^*$ at which a switch from a conventional junction to a $\pi$-junction takes place due to the competition between $K^>$ and $K^<$. We then consider the effect of finite temperature in Sec.~\ref{sec-finite}. Here analytical results can be obtained in the regime $T\lesssim T_c$.

\subsection{Weak coupling $\Gamma_S \ll \Delta_0$ at $T=0$}
\label{sec-weak}
		
At $T=0$, the gap in the leads is $\Delta=\Delta_0$. Let us first consider the contributions to the supercurrent from energies $E>\Delta_0$. In that regime, we find
		\begin{equation}
			\xi^R\approx \begin{cases}
				E,& E-\Delta_0\gg\frac{\Gamma_S^2}{\Delta_0},\\
				\left(\frac{\Gamma_S}{\sqrt{\Gamma_S^2-\Gamma_\phi^2}}+i\frac{\sqrt{\Gamma_S^2-\Gamma_\phi^2}}{\sqrt{2\Delta_0(E-\Delta_0)}}\right)\Delta_0,&E\to\Delta_0^+.
			\end{cases}
		\end{equation}
		Thus, as $E\to\Delta_0^+$, the spectral current saturates at $j(\Delta_0^+)=-\Gamma_S\Delta_0(\Gamma_S^2-\Gamma_\phi^2)^{-3/2}$, and $K^>$ can be approximated as
		\begin{equation}
		\label{eq:Ksupweak}
		K^>\approx-\int_{\Delta_0+{\Gamma_S^2}/{\Delta_0}}^\infty dE\;\frac1{E(E^2-\Delta_0^2)}\approx-\ln\frac{\Delta_0}{\Gamma_S}
		\end{equation}
		with logarithmic accuracy.
		(Note that the case $\Gamma_\phi\approx\Gamma_S$, which is realized when $\phi\approx 2\pi n$ would require more careful consideration. However, as the critical current is realized at phases $\phi\approx \pi(n+\frac12)$, we will not detail it here.)

Let us now turn to the subgap contributions. At weak coupling, the spectral gap $E_g$ is small. Namely,  Eq.~\eqref{eq-Eg} yields $E_g\approx \Gamma_{\phi}\ll\Delta_0$, varying between $E_g=|\Gamma_{S_1}-\Gamma_{S_2}|$ at $\phi=\pi n$ and $E_g=\Gamma_S$ at $\phi=\pi (n+1/2)$. Furthermore,
		\begin{equation}
			\xi^R\approx \begin{cases}
				\sqrt{E^2-E_g^2},& \Delta_0-E\gg\frac{\Gamma_S^2}{\Delta_0},\\
				\sqrt{\frac{(\Gamma_S^2-\Gamma_\phi^2)\Delta_0}{2(\Delta_0-E)}},&E\to\Delta_0^-.
			\end{cases}
		\end{equation}
		For $\Delta_0-V<\frac{\Gamma_S^2}{\Delta_0}$, we can thus approximate
		\begin{eqnarray}
			\label{eq:kn_k}
			K^<~&\approx&\int_{\max(E_g,eV)}^{\Delta_0-\Gamma_S^2/\Delta_0}dE\;\frac{\Delta_0^2}{\sqrt{E^2-E_g^2}(\Delta_0^2-E^2)}\\
			&\approx&\begin{cases}\ln\frac{\Delta_0^2}{\Gamma_SE_g},&eV<E_g,\\
				\ln\frac{\Delta_0\sqrt{\Delta_0^2-(eV)^2}}{\Gamma_SV},&eV\gg E_g.
			\end{cases}
			\nonumber
		\end{eqnarray}
		(For $\phi\to \pi n$,  $E_g$ has to be replaced by max$(\Gamma_\phi,\Gamma_N)$ in the above formulas.)
		
With Eqs.~\eqref{eq:Ksupweak} and \eqref{eq:kn_k}, we find the equilibrium ($V=0$) supercurrent 
		\begin{equation}
			I_1^{S,{\rm \, eq}}\approx\frac{\Gamma_S}{eR}\ln\frac{\Delta_0}{E_g}\;\sin\phi.
		\end{equation}
The result describes a conventional junction with current-phase relation $I^S(\phi)=I_c\sin\phi$ (neglecting the non-sinusoidal corrections due to phase-dependence of $E_g$ in the logarithm) and critical current $I_c^{\rm eq}=(\Gamma_S/eR)\ln(\Delta_0/E_g(\pi/2))$.

		Increasing the voltage in the subgap regime to $eV\gg E_g$ (still $eV<\Delta_0$), we obtain 
		\begin{equation}
			\label{eq-vstar}
			I_1^{S}(V)\approx\frac{\Gamma_S}{eR}\ln\frac{\sqrt{\Delta_0^2-(eV)^2}}{V}\;\sin\phi.
		\end{equation}
		The prefactor in that expression changes sign at $eV^*=\Delta_0/\sqrt{2}$, signaling the transition to a $\pi$-junction. Namely, at $V>V^*$, the current phase relation has the form  $I^S(\phi)=-I_c\sin\phi$ with critical current $I_c=(\Gamma_S/eR)\ln(V/\sqrt{\Delta_0^2/e^2-V^2})$.	
		
	Finally at $eV>\Delta_0$, the supercurrent saturates at
		\begin{equation}
			I_1^{S,\,>}\approx-\frac{\Gamma_S}{eR}\ln\frac{\Delta_0}{\Gamma_S}\;\sin\phi,
		\end{equation}
describing a  $\pi$-junction with critical current 	$I_c^>=(\Gamma_S/eR)\ln(\Delta_0/\Gamma_S)$ of the same order of magnitude as the equilibrium critical current.

Our results are in agreement with the original work by Volkov~\cite{volkov1995new}.
	
	\subsection{Strong coupling  $\Gamma_S \gg \Delta_0$ at $T=0$}
	\label{sec-strong}
		
As in the case of weak coupling, we start by considering the contributions to the supercurrent from energies $E>\Delta_0$. Here
		\begin{equation}	
			\xi^{R}\approx
			E+i\frac{\Gamma_SE}{\sqrt{E^2-\Delta^2}},
		\end{equation}
such that
		\begin{equation}
		\label{eq:Ksupstrong}
		K^>\approx-\Delta_0^2\int_{\Delta_0}^\infty dE\;\frac1{E(E^2+\Gamma_S^2)}\approx-\frac{\Delta_0^2}{\Gamma_S^2}\ln\frac{\Gamma_S}{\Delta_0}
		\end{equation}
		with logarithmic accuracy.

		Let us now turn to the subgap contributions. At strong coupling, $E_{g} \approx \Gamma_{\phi}\Delta_0/\Gamma_S$, varying  from $E_g^{\rm min}=|\Gamma_{S_1}-\Gamma_{S_2}|\Delta_0/\Gamma_S$ at $\phi=2\pi(n+1/2)$ to $E_g^{\rm max}=\Delta_0$ at $\phi=2\pi n$. Except for the narrow regime $\Gamma_S-\Gamma_\phi\ll\frac{\Delta_0^2}{\Gamma_S}$ corresponding to phases $|\phi-2\pi n|\ll\frac{\Delta_0}{\Gamma_S}$ (that is not relevant for determining the critical current, see below), we find	
		\begin{equation}	
			\xi^{R}\approx
			\Gamma_S\sqrt{\frac{E^2-E_g^2}{\Delta_0^2-E^2}},	
		\end{equation}
		and consequently
		\begin{widetext}

		\begin{equation}\label{eq:kless_approx}
			K^<\approx\frac{\Delta_0^2}{\Gamma_S}\int_{\max(E_g,eV)}^{\Delta_0}dE\;\frac{1}{\sqrt{E^2-E_g^2}\sqrt{\Delta_0^2-E^2}} =\begin{cases}
				\frac{\Delta_0}{\Gamma_S}K\left(1-\frac{E_g^2}{\Delta_0^2}\right), & eV<E_g,\\
				\frac{\Delta_0}{\Gamma_S}F\left(\arcsin\left(\sqrt{\frac{\Delta_0^2-(eV)^2}{\Delta_0^2-E_g^2}}\right)|1-\frac{E_g^2}{\Delta_0^2}\right), & E_g<eV<\Delta_0,
			\end{cases}
\end{equation}
				\end{widetext}
		where $K$ and $F$ are the complete and incomplete elliptic integrals of the first kind, respectively.
		( Similarly to the weak coupling case, for $\phi\to 2\pi (n+1/2)$, $E_g$ has to be replaced by max$(\Gamma_\phi,\Gamma_N)$ in the above formulas.)
		
		The equilibrium current-phase relation is given as
		\begin{equation}
			I_1^{S,{\rm \, eq}}\approx\frac{\Delta_0}{eR}\;K\left(\frac{4\Gamma_{S_1}\Gamma_{S_2}}{\Gamma_S^2}\sin^2\frac\phi2\right)\sin\phi.
		\end{equation}
		The critical current is $I_c^{\rm eq}\sim \Delta_0/(eR)$; it is realized at $\phi=\phi_c^{\rm eq}\in[0,\pi]$, and it corresponds to a conventional junction. (For $\Gamma_{S_1}=\Gamma_{S_2}$, one finds $\phi_c^{\rm eq}\approx0.59\pi$.)
		
		The fact that  $E_g$ reaches $\Delta_0$ (at  $\phi=2\pi n$) and that the contributions to the equilibrium supercurrent from energies outside the gap are  parametrically smaller than the contributions from subgap energies, as can be seen by comparing Eqs.~\eqref{eq:Ksupstrong} and \eqref{eq:kless_approx}, leads to a qualitatively different scenario for the current reversal  compared with the weak-coupling case.  The current-phase relation does not differ from the equilibrium case until $eV$ reaches $E_g^{\rm min}$. As the voltage further increases, the phase $\phi^*$ at which $E_g(\phi^*)=eV$ decreases from $\pi$ to $0$. Once it reaches $\phi_c^{\rm eq}$, the critical current starts to decrease (see  also Ref. \cite{bezuglyi2003}). Analyzing the phase dependence of the supercurrent around $E_g(\phi^*)=eV$, we conclude that $I_c(V)=I_1^{S,{\rm \, eq}}(\phi^*(V))$ for $eV\gtrsim E_g(\phi_c^{\rm eq})$. 
		
\begin{figure}
			\centering
			\includegraphics[width=\linewidth]{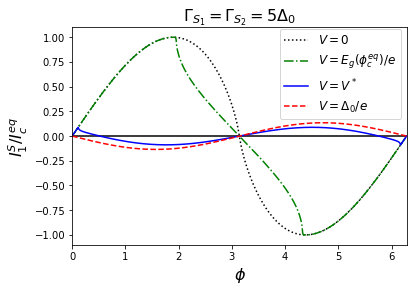}
			\caption{\label{fig:cur_phase_strong} The current-phase relation in the strong coupling case for different voltages. The equilibrium current-phase relation (dotted line) is non-sinusoidal. In the symmetric case, $\Gamma_{S_1}=\Gamma_{S_2}$,  shown here, the equilibrium critical current is realized at $\phi_c^{\rm eq}\approx0.59\pi$. The critical current starts to decrease once $eV=E_g(\phi_c^{\rm eq})$ (dash-dotted line). The transition takes place when the two extrema at positive and negative values of $I_1^S$ within the interval $[0,\pi]$  have the same magnitude (solid line). Thus the critical current at the transition is finite.
			For $eV>\Delta_0$, the current-phase relation takes the form $I_1^{S,>}=I_c^>\sin(\phi+\pi)$ (dashed line). }
		\end{figure}

		For $\Delta_0-eV \ll \Delta_0-E_g$,  we may approximate the incomplete elliptic integral of the first kind as
		$$F\left(\arcsin\left(\sqrt{\frac{\Delta_0^2-(eV)^2}{\Delta_0^2-E_g^2}}\right)|1-\frac{E_g^2}{\Delta_0^2}\right)\approx \sqrt{\frac{\Delta_0^2-(eV)^2}{\Delta_0^2-E_g^2}},$$
		which leads to a current-phase relation of the form
		\begin{equation}I_1^S(V)\approx \frac{\Delta_0}{eR}\left( \sqrt{\frac{\Delta_0^2-(eV)^2}{\Delta_0^2-E_g^2}}-\frac{\Delta_0}{\Gamma_S}\ln\frac{\Gamma_S}{\Delta_0}\right)\sin\phi.\label{eq:b}\end{equation}
		Thus the prefactor changes sign when $eV$ reaches $\sqrt{\Delta_0^2-(\Delta_0^2-E_g^2)\frac{\Delta_0}{\Gamma_S}\ln\frac{\Gamma_S}{\Delta_0}}$.
		In particular at the phase $\phi=\pi/2$, which gives the critical current at $V>V^*$, the sign changes when $\Delta_0-eV\approx \Delta_0^3/(4\Gamma_S^2)\ln^2(\Gamma_S/\Delta_0)$.
		A $\pi$-junction is realized once $|I_1^S(V,\pi/2)|$ exceeds $I_1^{S,{\rm \, eq}}(\phi^*(V))\approx(\pi\Delta_0/2eR)\phi^*(V)$ at voltages $\Delta_0-eV^*\approx\Delta_0^3/[2(1+\pi)^2\Gamma_S^2]\ln^2(\Gamma_S/\Delta_0)$. As a consequence, the critical current does not vanish at the transition. Such a behavior is characteristic of junctions with a non-sinusoidal current-phase relation.  Fig.~\ref{fig:cur_phase_strong} shows current-phase relations at different voltages to illustrate the above scenario.
		
		At $eV>\Delta_0$, the supercurrent saturates at
		\begin{equation}
		\label{eq:cur_greater_strong}	I_1^{S,\,>}=-\frac{\Delta_0^2}{eR\Gamma_S}\ln\frac{\Gamma_S}{\Delta_0}\;\sin\phi.
		\end{equation}
		Here the critical current $I_c^>=(\Delta_0^2/eR\Gamma_S)\ln({\Gamma_S}/{\Delta_0})$ is parametrically smaller than the equilibrium supercurrent $I_c^{\rm eq}$. 
		
		Note that, in the limit $\Gamma_S\to\infty$, we recover the result of Ref.~\cite{heikkila2002supercurrent} that the supercurrent gradually decreases with voltage and vanishes at $V>\Delta_0$.

	\subsection{Finite temperature}
	\label{sec-finite}
	
	As a next step, we consider the effects of finite temperature and, in particular, the regime $T\lesssim T_c$. In that regime, the equilibrium supercurrent is reduced due to the fact that the distribution function $f^{\rm eq}_L=\tanh( E/{2T})$ suppresses the low-energy contributions to the supercurrent. This also changes the competition between contributions from energies below and above the gap, and therefore affects the voltage $V^*$ at which the $\pi$-transition takes place. As, close to $T_c$, the gap is given as $\Delta/T_c \approx \sqrt{ 8\pi^2/{7\zeta(3)}} \sqrt{1-T/T_{c}}$, the relation $\Gamma_S\gg\Delta$ holds at arbitrary coupling.

As before, let us start by considering the contributions  to the supercurrent from energies outside the gap. We find
	\begin{eqnarray}
		K^>&\approx&-\Delta^2\int_\Delta^\infty dE\;\tanh\frac{E}{2T}\frac1{E(E^2+\Gamma_S^2)}\\
		&\approx&\begin{cases} -\frac{\pi\Delta^{2}}{4\Gamma_S T_{c}}+\frac{7\zeta(3)\Delta^{2}}{4\pi^{2}T_{c}^{2}},&\Gamma_S\ll T_c\sim\Delta_0,\\
			-\frac{\Delta^2}{\Gamma_S^2}\log\left(\frac{\Gamma_S}{T_c}\right),&\Gamma_S\gg T_c\sim\Delta_0.\end{cases}\nonumber
	\end{eqnarray}
	In weak coupling, the result is obtained using
	\begin{eqnarray*}
		\int_{\Delta/(2T_c)}^\infty dx\;\frac{\tanh x}{x(x^2+(\frac{\Gamma_S}{2T_c})^2)}&\approx&\int_0^\infty dx\;\frac1{x^2+(\frac{\Gamma_S}{2T_c})^2}\\
		&&+\int_0^\infty dx\;\frac{\tanh x-x}{x^3},
	\end{eqnarray*}
	where the second term has to be kept as, in equilibrium, the first term is canceled by the contributions from subgap energies (see below).
		
We now turn to the subgap contributions.	As $E\ll T\sim T_c$, the distribution function may be approximated as
	\begin{equation}f_L(V)
	\approx\frac E{2T_c}\frac1{\cosh^2({eV}/{2T_c})},\end{equation}
which yields
	\begin{eqnarray}
		K^<&\approx&\frac{\Delta^2/{2\Gamma_ST_c}}{\cosh^2({eV}/{2T_c})}\int_{E_g}^{\Delta}dE\;\frac{E}{\sqrt{E^2-E_g^2}\sqrt{\Delta^2-E^2}}\nonumber\\
		&=&\frac{\pi\Delta^2}{4\Gamma_ST_c}\frac1{\cosh^2({eV}/{2T_c})}.
	\end{eqnarray}
	As at $T=0$, the contributions from energies below and above the gap are comparable in the weak coupling case, $\Gamma_S\ll \Delta_0$.
	In that regime, the supercurrent is given as
	\begin{equation}
		I_1^S(V)\approx \frac{\Gamma_S}{eR}\frac{\Delta^{2}}{4\pi^2T_{c}^{2}}\left[7\zeta(3)-\pi^3\frac{T_c}{\Gamma_S}\tanh^{2}\frac{eV}{2T_{c}}\right]\sin\phi.
	\end{equation}
	As the dominant contributions cancel at $V=0$, i.e., as the critical current is parametrically smaller than the individual terms, the sign change happens at a small voltage,
	\begin{equation}
		eV^{*}\approx\frac{2}{\pi}\sqrt{\frac{7\zeta(3)\Gamma_ST_c}{\pi}}.
	\end{equation}		
	Interestingly the supercurrent at $V\gg V^*$ parametrically exceeds the equilibrium supercurrent, namely $I_c^{\rm eq}\sim\Gamma_S\Delta^2/(eRT_c^2)$,
	whereas
	$I_c^{>}\sim\Delta^2/(eRT_c)$. This enhancement occurs in the small temperature range $T_c-T\ll \Gamma_S^2/T_c$.

	By contrast, in the strong coupling case, $\Gamma_S\gg \Delta_0$, the subgap contributions dominate in equilibrium. Therefore a large voltage is needed to achieve the $\pi$-transition. In particular,
	\begin{equation}
		I_1^S(V)\approx\frac{\Gamma_S}{eR}\frac{\pi\Delta^{2}}{4\Gamma_S T_{c}}\left[\frac1{\cosh^{2}({eV}/{2T_{c}})}-\frac{4T_c}{\pi\Gamma_S}\ln\frac{\Gamma_S}{T_c}\right]\sin\phi,
	\end{equation}
	yielding	
	\begin{equation}
		eV^*\approx T_c \ln\left(\frac{\Gamma_S}{T_c }\right).
	\end{equation}
Here the critical current at large voltages, $I_c^>=\Delta^2/(eR\Gamma_S)$, is parametrically smaller than the equilibrium critical current, $I_c^{\rm eq}=\pi\Delta^2/(4eRT_c)$.

\subsection{Numerical results} 

	\begin{figure}
	\centering
	\begin{tabular}{cc}
		(a)\includegraphics[width=0.94\linewidth]{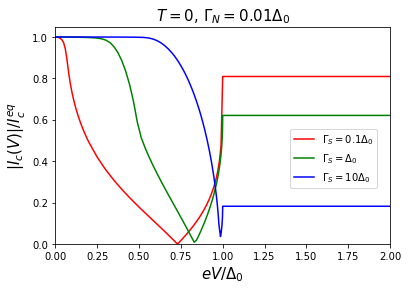}\\
		(b)\includegraphics[width=0.92\linewidth]{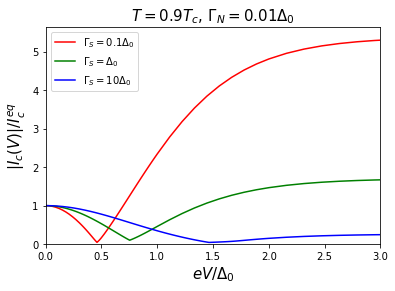}\\	
	\end{tabular}
\caption{\label{fig:critcur_v_no_inelastic} Critical current as a function of the voltage applied to the normal leads in the absence of inelastic relaxation in the dot. The minimum of the critical current at finite voltage signals the transition from a 0-junction to a nonequilibrium $\pi$-junction.  The critical currents are normalized by the (temperature-dependent) critical current in equilibrium. (a) Results for $T=0$ at different values of $\Gamma_S$. The critical current saturates at $V=\Delta_0$. (b) Results at $T=0.9T_c$. Interestingly, in the weak coupling limit, the  critical current at high voltages largely exceeds the equilibrium critical current.}
\end{figure}

\begin{figure}
		\centering
		\begin{tabular}{cc}
		(a)\includegraphics[width=0.92\linewidth]{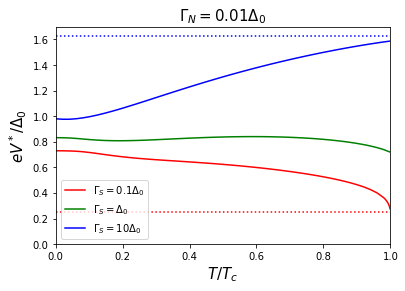}\\
			(b)\includegraphics[width=0.92\linewidth]{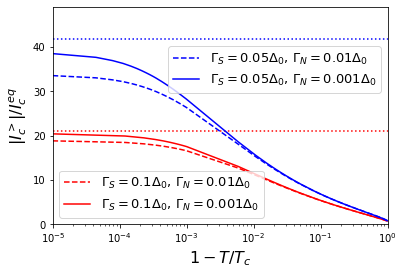}\\
					\end{tabular}
	\caption{\label{fig:finite_T} (a) Dependence of the voltage $V^*$, at which the transition from a  conventional junction to a nonequilibrium $\pi$-junction occurs, on temperature. The behavior is qualitatively different at weak coupling, where $V^*$ decreases with temperature, and strong coupling, where $V^*$ increases with temperature. The dotted lines show the analytical results that were obtained close to $T_c$. (b) Critical current enhancement in the $\pi$-phase in the weak coupling limit close to $T_c$. The dotted lines show the enhancement factor $I_c^>/I_c^{\rm eq}=4T_c/(\pi\Gamma_S)$ in the limit $\Gamma_N\to0$. A reduction at finite $\Gamma_N$ is observed.}
	\end{figure}

To visualize the results, we evaluate the critical current numerically. Figure~\ref{fig:critcur_v_no_inelastic} shows the voltage dependence of the critical current for various coupling strengths, at $T=0$ [Fig.~\ref{fig:critcur_v_no_inelastic}(a)] and at $T\lesssim T_c$ [Fig.~\ref{fig:critcur_v_no_inelastic}(b)]. The minimum of the critical current at a finite voltage signals the transition to a $\pi$-junction. The transition happens in all parameter regimes with the characteristic voltage $V^*$ increasing with coupling strength $\Gamma_S$. Numerically $V^*$ is obtained by determining the voltage at which the maximal current $I_{\rm max}(\phi)$ in the interval $\phi\in[0,\pi]$ changes sign. Figure~\ref{fig:finite_T}(a) shows the temperature dependence of $V^*$ for various coupling strengths. The critical current vanishes at the transition in the weak-coupling limit; it increases with $\Gamma_S$ as the current-phase relation becomes non-sinusoidal. The enhancement of the nonequilibrium critical current close to $T_c$ in the weak-coupling limit, as discussed in Sec.~\ref{sec-finite}, is illustrated in Fig.~\ref{fig:finite_T}(b). Here the dependence on $\Gamma_N$ is taken into account as well: the effect is seen to diminish as $\Gamma_N$ increases.

	\section{Effects of inelastic relaxation \label{sec:inelrel}}

 So far we neglected inelastic relaxation within the dot. Such relaxation processes, if strong enough, tend to establish a Fermi-Dirac distribution in the dot with an effective temperature and chemical potential determined by the coupling to the reservoirs. As the non-equilibrium $\pi$-junction relies on deviations from a Fermi-Dirac distribution in the dot, it is expected that strong enough inelastic processes will destroy the effect. Here we show that nevertheless the $\pi$-junction remains robust in a large regime of parameters.
	
	To determine the effect of internal relaxation, we have to compare it with the relaxation to the reservoirs. In the subgap regime, relaxation can take place only with the normal reservoirs. As we assume that $\Gamma_N$ is small, this is a very slow process and internal relaxation should start playing a role as soon as the corresponding rate exceeds $\Gamma_N$.
	
	A simple way to model internal relaxation is to couple the system to a fictitious fermionic bath~\cite{Buttiker1986} whose temperature and chemical potential are chosen such that the energy and charge currents between the dot and the bath vanish. We denote the temperature and chemical potential of this fictitious bath $T_b$ and $V_{b}$, respectively. The coupling between the dot and the bath is characterized by the rate $\Gamma_b$. The advantage of this description is that it is readily described using the general formulas in Sec.~\ref{sec:model}, extending the sum over normal leads to include the fictitious bath. We will assume $\Gamma_b\ll\Gamma_S$.
	
	The condition for the vanishing of the charge current between the dot and the bath can be deduced from Eq.~\eqref{eq:cur_N}, namely
	\begin{equation}
		\int dE\;(f_T-f_{Tb}){\rm Re}(\cos\theta)=0.
	\end{equation}
	The energy current $J$ can be written as
	\begin{equation}
		J_{b}=\frac{G_b}{16e}\int dE\; E\tilde{J}_{b}(E)
	\end{equation}
	with
	\begin{equation}
		\tilde{J}_{b}=\text{Tr}\left(\hat{g}_b^K\hat{g}^A-\hat{g}^R\hat{g}_b^K-\hat{g}^K\hat{g}_b^A+\hat{g}_b^R\hat{g}^K\right),
	\end{equation}
	which, using similar considerations as the ones leading to Eq.~\eqref{eq:cur_N}, yields the condition
	\begin{equation}
		\int dE\;E(f_L-f_{Lb}){\rm Re}(\cos\theta)=0.
	\end{equation}
	Using the expressions for the distribution functions in the dot, Eqs.~\eqref{eq:fl} and \eqref{eq:ft}, the two conditions can be rearranged such that the left-hand side only depends on the parameters $T_b$ and $V_b$ of the fictitious bath, whereas the right-hand side only depends on the parameters $T$ and $V_n$ of the normal reservoirs. For simplicity, we will consider only the case $T=0$ here. For the setup considered in the previous section, this yields
	\begin{widetext}
	\begin{eqnarray}
		\int_0^\infty dE\;f_{Tb}\left(1-\frac{\Gamma_b}{C\,\text{Im}\,\xi^R}\right){\rm Re}(\cos\theta)&=&(\Gamma_{N_<}-\Gamma_{N_>})\int_0^{eV}dE\;\frac{1}{C\,\text{Im}\,\xi^R}{\rm Re}(\cos\theta),\label{eq:vbtb1}\\
		\!\!\!\!\!\!\int_0^\infty dE\;E(1-f_{Lb})\left(1-\frac{C\Gamma_b}{\text{Im}\,\xi^R}\right){\rm Re}(\cos\theta)&=& \Gamma_N\int_0^{eV}dE\;E\frac{C}{\text{Im}\,\xi^R}{\rm Re}(\cos\theta),\label{eq:vbtb2}
	\end{eqnarray}
	\end{widetext}
	where the parameters $\xi^R$ and $C$ are specified in Sec.~\ref{sec:model}.
	
	In a symmetric setup, $\Gamma_{N_>}=\Gamma_{N_<}$, the right-hand side of Eq.~\eqref{eq:vbtb1} vanishes. This imposes $V_b=0$ such that $f_{Tb}=0$.  We will concentrate on this case to illustrate the effect of relaxation. Extensions to an asymmetric case are discussed in Appendix \ref{ap_gmminus}. We find that the characteristic voltage $V^*$ of the $\pi$-transition depends on the asymmetry only very weakly. Thus the following results are  qualitatively valid also in the extreme case of only one normal lead.

\begin{figure}
		\centering
		\begin{tabular}{cc}
			(a)\includegraphics[width=0.94\linewidth]{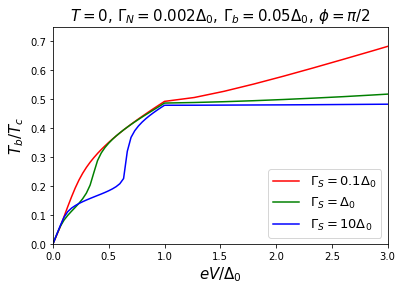}\\
			(b)\includegraphics[width=0.94\linewidth]{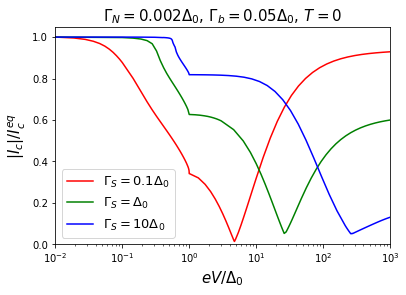}
			\end{tabular}
		\caption{\label{fig:tb_critcur_v} (a) Dependence of the bath temperature on voltage in a symmetric setup, obtained by numerically solving Eq.~\eqref{eq:vbtb2}. As discussed in the text, the temperature at $eV=\Delta_0$ is found to be almost independent of the coupling strength to the superconducting leads. Note that the subgap temperature-dependence varies with phase due to the phase-dependence of $E_g$. (b) Critical current as a function of voltage (log-scale) for the same parameters as in (a), obtained numerically from Eq.~\eqref{eq:inel_current}. Inelastic relaxation weakens the nonequilibrium effects and pushes the $\pi$-transition to large voltages.}
		\end{figure}

In the absence of superconductivity,	the bath temperature $T_b$ is readily obtained from Eq.~\eqref{eq:vbtb2} setting $\Delta_0=0$. In that case, $T_b=\sqrt{3\Gamma_N/\pi^2(\Gamma_N+\Gamma_S)}eV$. Superconductivity suppresses relaxation to the superconducting leads at low energies. This leads to a faster rise of
the temperature in the subgap regime. 
The temperature obtained by solving Eq.~\eqref{eq:vbtb2} numerically is shown in Fig.~\ref{fig:tb_critcur_v}(a), where we consider the case $\Gamma_N \ll \Gamma_b \ll \Gamma_S$ for different strengths of $\Gamma_S$. The results can be understood qualitatively as follows:
At small $V$, the S leads play no role in the heat balance process. They do, however, modify the density of states in the dot. Initially only states at energies $E<E_g$ are accessible in the dot. (Their density of states is finite when taking into account finite $\Gamma_N$ and $\Gamma_b$.) In that regime, $T_b=\sqrt3 eV/\pi$. As temperature increases, the integral on the left hand side of Eq.~\eqref{eq:vbtb2} will contain contributions $\sim e^{-T_b/E_g}$ from energies $E>E_g$. Due to the increased density of states at $E>E_g$, their contribution can be shown to become relevant at $T_b\sim E_g/\ln(\Gamma_S/\Gamma_b)\ll E_g$ and to lead to a slow-down of the increase in temperature.
Once $eV>E_g$, the enhanced density of states becomes accessible in the integral on the right hand side of Eq.~\eqref{eq:vbtb2}. This leads to an enhanced power injection and results in a sharp increase in $T_b$. At $eV=\Delta_0$, we can approximate Eq.~\eqref{eq:vbtb2} as
	\begin{eqnarray}
		\frac{\Gamma_N}{\Gamma_b}\int_0^{\Delta_0} dE\;E f_{Lb}\approx \int_{\Delta_0}^\infty dE\;E(1-f_{Lb}),
			\end{eqnarray}
yielding $T_b\sim\Delta_0/\ln(\Gamma_b/\Gamma_N)$, i.e., a temperature that is almost independent of the coupling to the superconducting leads. Finally at $eV>\Delta_0$, the slope is determined by the normal state result $T_b=\sqrt{3\Gamma_N/(\Gamma_N+\Gamma_S)}eV/\pi$.

	Once $T_b$ is determined, the supercurrent can be computed from Eq.~\eqref{supcur_gen_exp}. For the specific case considered here and $\Gamma_N,\Gamma_b\ll\Gamma_S$, it takes the form
	\begin{equation}
		\label{eq:inel_current}I^S=\frac{\Gamma_S}{eR}\sin\phi \left[K^>+\tilde K^<(T_b)\right].
	\end{equation}
	with 
	\begin{eqnarray}
	\label{eq-distrelax}
	\tilde K^<(T_b)&=&\Theta(\Delta_0-eV)\frac{\Gamma_N}{\Gamma_N+\Gamma_b}\int_{eV}^{\Delta_0}\!\!\!dE\;j(E)\nonumber\\&&+\frac{\Gamma_b}{\Gamma_N+\Gamma_b}\int_0^{\Delta_0}\!\!\!dE\;f_{Lb}(E)j(E).
	\end{eqnarray}
If $\Gamma_b\ll\Gamma_N$,  $\tilde K^<(T_b)=K^<$ up to  corrections of order $\Gamma_b/\Gamma_N$.

As the corrections to $K^<$ are positive, they tend to increase $V^*$. As they are small, one may expect that the corrections to $V^*$ are small. This turns out to be true only as long as $eV^*$ remains smaller than $\Delta_0$. Once $eV^*$ reaches $\Delta_0$, it increases rapidly upon further increasing $\Gamma_b$ due to the energy exchange with the superconducting leads. As seen in Sec.~\ref{sec-strong}, the transition happens very close to $\Delta_0$ in the strong coupling regime such that small corrections are sufficient to push $eV^*$ up to $\Delta_0$. Using Eq.~\eqref{eq:b}, we can estimate that this happens at $\Gamma_b \sim (\Gamma_N \Delta_0/\Gamma_S)\ln(\Gamma_S/\Delta_0)\ll\Gamma_N$.

In the following, we will concentrate on the opposite regime, $\Gamma_b\gg\Gamma_N$, where
	\begin{eqnarray}
	\tilde K^<(T_b)&\approx&\int_0^{\Delta_0}\!\!\!dE\;f_{Lb}(E)j(E)\nonumber\\&=&\int_{E_g}^{\Delta_0} dE\;\frac1{\xi^R}\frac{\Delta_0^2}{\Delta_0^2-E^2}\tanh\frac E{2T_b},
	\end{eqnarray}
corresponding to the equilibrium result, but at finite temperature $T_b$.

\begin{figure}
	\includegraphics[scale=.6]{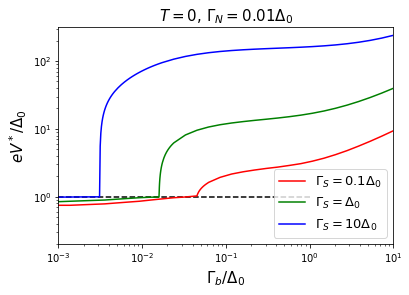}
	\caption{\label{fig:vst_gamma_b} The effect of inelastic relaxation in the dot, characterized by the rate $\Gamma_b$ on the voltage $V^*$, at which the $\pi$-transition occurs, for the case $T=0$. $eV^*$ is found to increase slowly until it reaches $\Delta_0$ followed by a much faster variation, if $\Gamma_ b$ is further increased. At $\Gamma_b\gg\Gamma_N$, the system is always in the latter regime. }
\end{figure}

As long as $T_b\ll E_g$ the effect of the finite temperature is negligible. Analytic results can be obtained for $E_g\ll T_b\ll\Delta_0$ (relevant for $\Gamma_S\ll\Delta_0$ only), where
\begin{equation}
	\tilde K^<(T_b)\approx\Delta_0^2\int_{T_b}^{\Delta_0-\Gamma_S^2/\Delta_0}dE\;\frac{1}{E(\Delta_0^2-E^2)}\approx\ln\frac{\Delta_0^2}{\Gamma_ST_b},
\end{equation}
as well as for $T_b\gg\Delta_0$, where
\begin{equation}
	\label{eq:Tb-large1}
	\tilde K^<(T_b)\approx \frac{\Delta_0^2}{2T_b}\int_{0}^{\Delta_0-\Gamma_S^2/\Delta_0}\;\frac{dE}{\Delta_0^2-E^2}\approx\frac{\Delta_0}{2T_b}\ln\frac{\Delta_0}{\Gamma_S}
\end{equation}
if $\Gamma_S\ll\Delta_0$ and
\begin{equation}
	\label{eq:Tb-large2}
		\tilde K^<(T_b)\approx \frac{\Delta_0^2}{2\Gamma_ST_b}\int_{0}^{\Delta_0}\;\frac{dE}{\sqrt{\Delta_0^2-E^2}}\approx\frac{\pi\Delta_0^2}{4\Gamma_ST_b}
\end{equation}		
if $\Gamma_S\gg\Delta_0$.
To find the $\pi$-transition, we have to compare these results with $K^>$ computed in the previous section, namely Eq.~\eqref{eq:Ksupweak} at weak coupling and Eq.~\eqref{eq:Ksupstrong} at strong coupling.
We see that in the weak coupling limit, $\Gamma_S\ll\Delta_0$, the transition happens when $T_b^*$ is of the order of $\Delta_0$. On the other hand, in the strong coupling limit, $\Gamma_S\gg\Delta_0$, the transition happens at $T_b^*\sim\Gamma_S/\ln(\Gamma_S/\Delta_0)\gg\Delta_0$. 
In both cases, the corresponding voltage $V^*$ is larger than $\Delta_0/e$, such that we may use the relation $eV^*\sim \sqrt{\frac{\Gamma_S}{\Gamma_N}}T_b^*$. At weak coupling, this yields
\begin{equation}
eV^*\sim \sqrt{\frac{\Gamma_S}{\Gamma_N}}\Delta_0,
\end{equation}
whereas at strong coupling we find
\begin{equation}
eV^*\sim\frac{\Gamma_S^{3/2}}{\sqrt{\Gamma_N}\ln\frac{\Gamma_S}{\Delta_0}}.
\end{equation}
Thus, in both cases, the transition from a  conventional junction to a nonequilibrium $\pi$-junction still occurs, though it is pushed to voltages $eV^*\gg\Delta_0$. Figure~\ref{fig:vst_gamma_b} shows $V^*$ as a function of $\Gamma_b$ for different strengths of $\Gamma_S$.  The rapid increase in $V^*$ once it has reached $\Delta_0$ is clearly seen.

	\section{Discussion and outlook \label{sec:concl}}

We showed that the supercurrent through a superconductor-normal dot-superconductor junction is strongly modified in the presence of quasiparticle injection via normal leads. The effect is most pronounced in the absence of inelastic relaxation in the dot, when the dot distribution function is very far from a Fermi-Dirac distribution. In the weak coupling case, the critical current may be suppressed to zero due a tiny quasiparticle injection current $I_N\sim \Gamma_N/{\rm min}\{\Gamma_S,\Delta_0\}I_c^{\rm eq}\ll I_c^{\rm eq}$ at moderate voltage $eV<\Delta_0$. Further increasing $V$ leads to a revival of the supercurrent, though with an inverted sign of the current-phase relation, corresponding to a $\pi$-junction. In the strong coupling case, a similar $\pi$-transition is observed, but due to the non-sinusoidal current-phase relation a finite critical current remains at the transition. The origin of this $\pi$-transition can be easily understood in the short junction setup considered here: the supercurrent is determined by a competition between subgap processes and processes involving energies outside the gap with opposite sign. Interestingly we find that, in the weak-coupling case, the critical current at high voltages deep in $\pi$-junction regime may largely exceed the equilibrium critical current  close to $T_c$.
 It is straightforward to generalize the results to multiterminal junctions as discussed in Appendix \ref{sec:cch}.

Internal relaxation in the dot leads to a more Fermi-Dirac like distribution function. This slows down the suppression of the critical current with increasing injection voltage. As long as the internal relaxation rate $\Gamma_b\ll\Gamma_S$, the $\pi$-transition is robust, but it occurs at much larger voltage. We expect the transition to be completely suppressed at $\Gamma_b\gg\Gamma_S$. Furthermore, in our study, we did not consider relaxation by phonons -- the only external relaxation processes are due to the currents to the leads. This requires the rate $\Gamma_N$ to be not too small. If the main external relaxation process is due to phonons, we also expect the $\pi$-transition to be absent. A suppression of the critical current due to the quasiparticle injection, the weaker the larger the phonon relaxation rate $\Gamma_{\rm ph}$, should remain.

The effect of quasiparticle injection on the critical current in a variety of setups has been intensively studied in recent years. Here we see in detail how a tiny quasiparticle injection current may completely modify the system properties in a very simple setup. Our study further highlights the importance of the shape of the quasiparticle distribution function with much stronger effects for a non Fermi-Dirac shape.

\acknowledgements
We thank T.~Jalabert and C.~Chapelier for useful discussions. We acknowledge support from the ANR through Grant No. ANR-16-CE30-0019.

\appendix
	
	\section{Derivation of some identities  for solving the kinetic equation \label{ap_relation}}
	
	To obtain Eq.~\eqref{eq:Keldysh3} for the Keldysh component of the dot Green function, we use the identity
\begin{equation}
		\label{eq:id1}
		\hat{\tau}_3-\hat{g}^{R}\hat{\tau}_3\hat{g}^{A}=C(\hat{g}^{R}-\hat{g}^{A}),
	\end{equation}
	with $C$ given by Eq.~\eqref{eq:id2}. To derive this identity, we start from the following parameterization of $\hat{g}^{R/A}$,
	\begin{eqnarray}
		\hat{g}^{R}&=&\sin\theta(\sin\chi\hat{\tau}_{1}+\cos\chi\hat{\tau}_{2})+\cos\theta \hat{\tau}_{3},\\
		\hat{g}^{A}&=&\sin\theta^{*} (\sin\chi\hat{\tau}_{1}+\cos\chi\hat{\tau}_{2})-\cos\theta^{*} \hat{\tau}_{3},
	\end{eqnarray}
	where we used that $\chi$ is real.
	Using trigonometric identities, we easily obtain
	\begin{widetext}
	\begin{eqnarray}
		\hat{g}^{R}-\hat{g}^{A}&=&2\cos\frac{\theta+\theta^{*}}{2}\left[\sin\frac{\theta-\theta^{*}}{2}(\sin\chi\hat{\tau}_{1}+\cos\chi\hat{\tau}_{2})+\cos\frac{\theta-\theta^{*}}{2}\hat{\tau}_{3}\right],\\
		\hat{\tau}_{3}-\hat{g}^{R}\hat{\tau}_{3}\hat{g}^{A}&=&2\cos\frac{\theta-\theta^{*}}{2}\left[\sin\frac{\theta-\theta^{*}}{2}(\sin\chi\hat{\tau}_{1}+\cos\chi\hat{\tau}_{2})+\cos\frac{\theta-\theta^{*}}{2}\hat{\tau}_{3}\right].
	\end{eqnarray}
	\end{widetext}
Thus, Eq.~\eqref{eq:id1} holds with
		\begin{equation}
		C=\frac{\cos\frac{\theta - \theta^*}2}{\cos\frac{\theta+ \theta^*}2}=\frac{1+\cos\theta\cos\theta^*+\sin\theta\sin \theta^*}{\cos\theta+\cos\theta^*}.
	\end{equation}
The further identity $\hat{g}^{R}\hat{\tau}_3-\hat{\tau}_3\hat{g}^{A}=C(\hat\tau_0-\hat{g}^{R}\hat{g}^{A})$ follows trivially from the normalization condition $\hat g^2=1$.

For the evaluation of the current, it is useful to show that $C$ simplifies for $|E|<\Delta$, resulting in a simple expression for $f_L$ in the dot as given in Eq. \eqref{eq:fL_simple}. 		For $|E|<\Delta$, $\cos\theta_{S}$ is purely imaginary and $\sin\theta_{S}$ is purely real, such that, using Eqs.~\eqref{eq:theta_chi} and
		\eqref{eq:gmphi_xir}, we can write
	\begin{equation}
		\cos\theta=\frac{a+i\Gamma_N}{\xi^R} \quad \text{and}\quad \sin\theta=\frac{ib}{\xi^R},
	\end{equation}
	with $\xi^{R}=\sqrt{(a+i\Gamma_{N})^{2}-b^{2}}$,
	where $a$ and $b$ are real numbers. As a consequence,
		\begin{equation}
		C=\frac{|\xi^R|^2+a^2+\Gamma_N^2+b^2}{2a{\rm Re}\,\xi^R+2\Gamma_N{\rm Im}\,\xi^R}.
		\label{eq:Cprelim}
	\end{equation}
Using ${\rm Re}\,\xi^R{\rm Im}\,\xi^R=a\Gamma_N$ and $({\rm Im}\,\xi^R)^2=-(a^2-\Gamma_N^2-b^2-|\xi^R|^2)/2$, it can easily be shown that Eq.~\eqref{eq:Cprelim} reduces to $C={\rm Im}\,\xi^R/\Gamma_N$.

			\section{Critical current hypersurfaces in multiterminal junctions \label{sec:cch}}
			
			\begin{figure}
				\centering
				\begin{tabular}{cc}
					\includegraphics[width=0.94\linewidth]{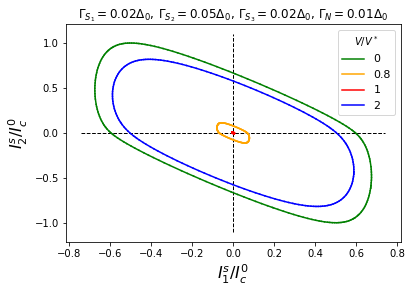}\\
					\includegraphics[width=0.94\linewidth]{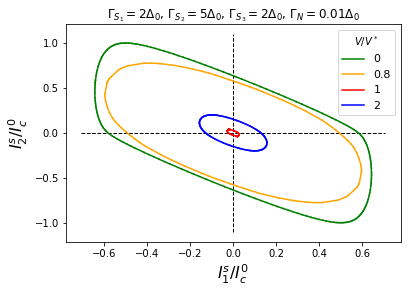}
				\end{tabular}
				\caption{\label{fig:ccc} Critical current hypersurfaces (CCH) for a setup with three superconducting leads at $T = 0$. The overall shape of the CCH depends on the values of all the couplings $\Gamma_s$. (a) Weak coupling. At $V^*$, the CCH shrinks to a point. (b) Strong coupling. At $V^*$, the CCH remains finite. The slight variations of $V^*$ for different leads are not visible on this scale. }
			\end{figure}
			
The general formula \eqref{supcur_gen_exp} can easily be applied to multiterminal junctions with more than two superconducting leads. In that case, there is not a single critical current, but one can define a critical current hypersurface (CCH)~\cite{pankratova2020multiterminal} in the space of $m-1$ independent supercurrents,  where $m$ is the number of superconducting leads. (The remaining current is determined by current conservation.)  The CCH encloses the hypervolume, where a non-dissipative supercurrent can flow.
According to Eq.~\eqref{supcur_gen_exp}, the supercurrents are given as
	\begin{equation}
		I_{s}^{S}=\frac1e\sum_{s'}G_{s}\Gamma_{s'}\sin\phi_{s's}\left(K^<+K^>\right).
	\end{equation}
In addition to the explicit phase dependence, $K^<$ depends on the phase difference through $E_g\propto\Gamma_\phi=\sqrt{\sum_{s,s'}\Gamma_{s}\Gamma_{s'}\cos\phi_{ss'}}$.

The same competition between $K^<$ and $K^>$ that was responsible for the $\pi$-transition in the two-terminal setup will lead to a non-monotonous dependence of the hypervolume enclosed by the CCH as a function of the voltage applied to the normal leads.

In the weak coupling case, $\Gamma_S\ll\Delta_0$, we saw that the phase-dependence of $K^<$  does not play an important role. Thus, the CCH will evolve with increasing voltage without changing its shape and shrink to a point at $V^*$ before increasing again. Here $V^*$ has the same value as for the two-terminal case, $eV^*=\Delta_0/\sqrt2$. On the other hand, in the strong coupling case, $\Gamma_S\gg\Delta_0$, the phase-dependence of $K^<$  does play an important role. Thus, the shape of the CCH will depend on voltage. Furthermore, as the critical currents never vanish, the CCH does not shrink to a point: it reaches a minimum at  $V^*$ before increasing again. As $V^*$ depends on the non-sinusoidal shape of the current-phase characteristic, one obtains the same order of magnitude as for the two-terminal case, $eV^*\lesssim\Delta_0$, but the minima for the critical currents corresponding to different leads may happen at slightly different voltages. 

Figure~\ref{fig:ccc} shows some examples of critical current hypersurfaces in a setup with three superconducting leads. They were obtained by evaluating the supercurrents using Eq.~\eqref{supcur_gen_exp} on a grid of $m-1$ independent phases taking values in the interval $[-\pi,\pi]$.

 \begin{figure}[h]
	\centering
	\begin{tabular}{cc}
		(a)\includegraphics[width=0.94\linewidth]{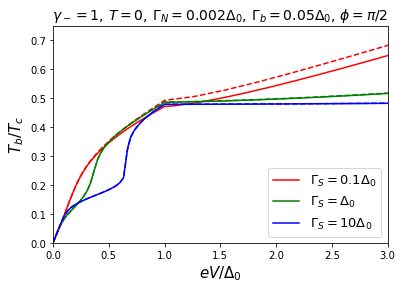}\\
		(b)\includegraphics[width=0.94\linewidth]{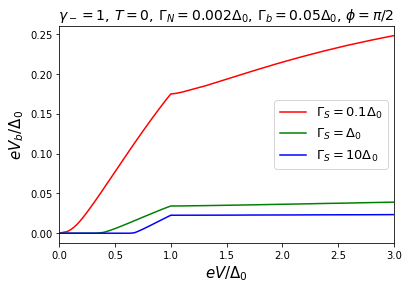}
	\end{tabular}
	\caption{\label{fig:vb_tb_asymmetric}  Dependence of (a) the bath temperature $T_b$ and (b) the bath chemical potential $V_b$ on voltage in an extreme asymmetric setup with $\gamma_-=1$. The same parameters as in Fig.~\ref{fig:tb_critcur_v} were used. The results for $T_b$ at $\gamma_-=0$ are shown as dashed lines for comparison.}
\end{figure}

 \begin{figure}[h]
	\centering
	\begin{tabular}{cc}
		(a)\includegraphics[width=0.94\linewidth]{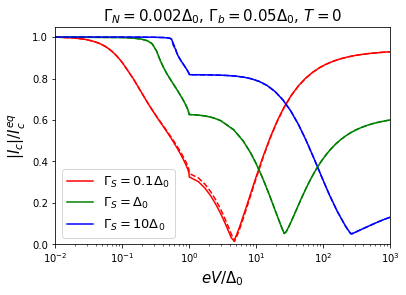}\\
		(b)\includegraphics[width=0.94\linewidth]{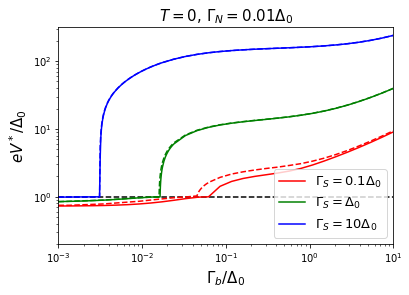}
	\end{tabular}
	\caption{\label{fig:ic_vst_asymmetric} (a) Critical current as a function of voltage in an extreme asymmetric setup with $\gamma_-=1$. The same parameters as in Fig.~\ref{fig:tb_critcur_v} were used. The results for $T_b$ at $\gamma_-=0$ are shown as dashed lines for comparison. (b) Critical voltage $V^*$ as a function of $\Gamma_b$ at $\gamma_-=1$. The results for $\gamma_-=0$ (Fig.~\ref{fig:vst_gamma_b}) are shown as dashed lines for comparison. Deviations can be seen in the weak coupling case at intermediate $\Gamma_b$.
	 }
\end{figure}

	\section{Inelastic relaxation in an asymmetric setup \label{ap_gmminus}}
	
In the main text, we considered the effect of inelastic relaxation only in a symmetric setup  such that the voltage $V_b$ of the fictitious bath remained 0. Here we address the question what happens in an asymmetric setup, where $\gamma_-=(\Gamma_{N_>}-\Gamma_{N_<})/\Gamma_N\neq0$. In that case, the subgap contributions to supercurrent depend on the distribution function
\begin{eqnarray}
f_{Lb}(E)=\frac12\left[\tanh\frac{E-eV_b}{2T_b}+\tanh\frac{E+eV_b}{2T_b}\right],
\end{eqnarray}
as can be seen from Eq.~\eqref{eq-distrelax}. Both $T_b$ and $V_b$ are expected to increase with voltage and gradually suppress the low-energy contributions to the supercurrent, eventually leading to a $\pi$-transition. To get an idea about their magnitude, let us first discuss the normal case, such that Eqs.~\eqref{eq:vbtb1} and \eqref{eq:vbtb2} simplify to
	\begin{eqnarray}
		\label{eq:app_vbtb_normal}
		&eV_b=\gamma_-\frac{\Gamma_N}{\Gamma_S}V,&\\
		&T_b^2\left[-{\rm Li}_2(-e^{eV_b/T_b})-{\rm Li}_2(-e^{-eV_b/T_b})\right]=\frac{\Gamma_N}{\Gamma_S}\frac{V^2}{2},\qquad&
	\end{eqnarray}
where $\rm{Li}_2$ is the dilogarithm function and we assumed $\Gamma_N \ll\Gamma_S$ as for the main part of this paper. This assumption ensures that $T_b\gg eV_b$, such that we can approximate
\begin{equation}
T_b\approx  \sqrt{\frac{3\Gamma_N}{\pi^2\Gamma_S}}eV\left(1-\frac{\pi^2\gamma_-^2}{6}\frac{\Gamma_N}{\Gamma_S}\right).
\end{equation}
We conclude that the finite $\gamma_-$ only leads to small  modifications of the distribution function compared to the symmetric case, corresponding to a shift of the temperature of the order $\delta T_b/T_b\sim \Gamma_N/\Gamma_S$.

At $\Gamma_b \gg\Gamma_N$,  the $\pi$-transition at $\gamma_-=0$ happens in the regime where $T_b$ is given by the normal state result. Thus the above considerations are sufficient to conclude that a finite $\gamma_-$ has negligible effect. This is further illustrated in Figures~\eqref{fig:vb_tb_asymmetric} and ~\eqref{fig:ic_vst_asymmetric}.


	\end{document}